\documentclass[prl,aps,floats,showpacs,amsmath,amssymb,preprint]{revtex4}
\usepackage{bm,graphicx,epsf}
\usepackage{bm,graphicx}
\newcommand{\be}{\begin{equation}}
\newcommand{\ee}{\end{equation}}   
\newcommand{\bea}{\begin{eqnarray}}
\newcommand{\eea}{\end{eqnarray}}
\newcommand{\phrl}[1]{Phys.~Rev.~Lett. {\bf #1}}
\newcommand{\phrb}[1]{Phys.~Rev.~B {\bf #1}}

\newcommand{\jpsj}[1]{J.~Phys.~Soc.~Jpn.{\bf #1}}
\newcommand{\lb}{\left[}
\newcommand{\rb}{\right]}
\newcommand{\lp}{\left(}
\newcommand{\rp}{\right)}

\newcommand{\q}{{\bf q}}
\renewcommand{\k}{{\bf k}}

\newcommand{\F}{{\cal F}}
\newcommand{\bib}{\bibitem}

\begin{document}

\title{ Derivation of the superconducting gap equation for the noncentrosymmetric superconductor $Li_2Pt_3B$}
\author{Soumya P. Mukherjee}
\affiliation{Asia Pacific Center for Theoretical Physics, Hogil Kim Memorial building 5th floor, POSTECH, Hyoja-dong, Namgu, Pohang 790-784, Republic of Korea} 
\author{Tetsuya Takimoto} 
\affiliation{Asia Pacific Center for Theoretical Physics, Hogil Kim Memorial building 5th floor, POSTECH, Hyoja-dong, Namgu, Pohang 790-784, Republic of Korea} 
\affiliation{Department of Physics, POSTECH, Pohang,Gyeongbuk 790-784, Republic of Korea} 

\date{\today}
\pacs{74.20.Mn,74.20.Rp,74.70.-b, 74.90.+n}

\begin{abstract}
We present here the mathematical background of our approach, presented in \cite{Soumya} regarding the gap function and symmetry for the noncentrosymmetric (NCS) superconductor $Li_2Pt_3B$. As revealed by the 
experiment, this NCS superconductor gives rise to line nodes in the superconducting order parameter, which is responsible for many of its experimental behaviors. Owing to the enhanced d-character of the 
relevant bands that cross the Fermi level,the system gets weakly correlated. The nature and symmetry of this nodal behavior is explained from a microscopic viewpoint. In this article starting with an Hubbard 
model relevant for this NCS system by considering the effect of the onsite Coulomb repulsion on the pairing potential perturbatively, we extract the superconducting gap equation. Further analysis \cite{Soumya} of 
this equation predicts a $s_\pm$ wave gap function with line nodes as the most promising candidate in the superconducting state. 

\end{abstract}

\maketitle

\section{Introduction}

  The discovery of superconductivity in the noncentrosymmetric compound $CePt_3Si$ \cite{Bauer} opened a new field of research. Broken inversion symmetry gives rise to an antisymmetric spin-orbit coupling 
which splits the spin degenerate electronic bands into two, each of which are now described by a helicity index. Gorkov and Rashba \cite{Rashba} already predicted that the superconducting order parameter in 
such a system is bound to be a mixture of spin-singlet as well as spin-triplet components. On the other hand, this gives rise to many unusual behaviors which are revealed through various experiments. Many 
possible theoretical explanations came up to explain these behaviors. For $CePt_3Si$ the fact that the superconducting phase coexists with the magnetic phase \cite{Bauer} makes it difficult to clarify the 
mechanism. 

The experimental finding of superconductivity in $Li_2Pd_3B$ \cite{Togano} and subsequently the experiments in the pseudo-binary complete solid solution $Li_2(Pd_{1-x}Pt_x)_3B$, $x=0\sim1$ 
\cite{Badica} revealed the fact that the superconducting phase for this family of compound does not coexists with other phases. Superconductivity in $Li_2Pd_3B$ is confirmed to be of s-wave type
\cite{Nishiyama1,Nishiyama2,Takeya1,Takeya2,Yuan,Yokoya}  and phonon mediated. On the other hand the compound $Li_2Pt_3B$, although iso-structural with $Li_2Pd_3B$, shows line node behavior in several 
experiments\cite{Nishiyama2,Takeya2,Yuan}.

It was shown by the band structure calculations of the compound $Li_2Pt_3B$ that the Fermi surface is nested and there is an enhancement of the d-character of the bands that cross the Fermi level 
\cite{Chandra,Lee}. This enhanced d-character gives rise to a correlation effect and we describe the Hamiltonian for this system \cite{Soumya} by adding an extra Hubbard like term with the usual Hamiltonian 
for the NCS metal,

\be
H=\sum_{\k \sigma \sigma'} \lp \lb \varepsilon_{\k}-\mu\rb \hat{\sigma}_{o}+ {\bf g}_{\k}.\hat{\sigma}\rp_{\sigma \sigma'} c^\dagger_{\k \sigma} c_{\k \sigma'} + U\sum_{i} n_{i \uparrow}n_{i \downarrow}
\label{H}
\ee

Here the symbols have their usual meaning. The band dispersion $\varepsilon_\k$ is obtained by a tight-binding fitting \cite{Soumya} and ${\bf g}_{\k}$ is the antisymmetric spin-orbit (ASO) coupling term 
specific to the point group of the system and is given by ${\bf g}_{\k}=g(\sin(k_x),\sin(k_y),\sin(k_z))$ . U describe the onsite Coulomb interaction as usually.

Our aim in this work is to describe the pathway to derive the superconducting gap equation [Eq.14 in \cite{Soumya}] for this cubic NCS superconductor starting from Eq.(\ref{H}). For this we first define a 
few relevant quantities. Firstly we define the dynamical susceptibility tensor as,
\be
\chi_{\alpha \beta}(\q,i\Omega_n)=\int^{1/T}_0{d\tau e^{i \Omega_{n}\tau}\langle T_\tau [S^\alpha_\q(\tau)S^\beta_{-\q}(0)] \rangle}
\label{sus}
\ee
here $\langle ...\rangle$ denotes the thermal average, $T_\tau$ is the imaginary time ordering and $\Omega_n$ are the Bosonic Matsubara frequencies. The charge ($\alpha=c$) and $\alpha$ component of the spin 
operators with wave vector $\q$ are defined as,
\be
S^c_{\q}=\frac{1}{2}\sum_{\k \sigma} c^\dagger_{\k \sigma} c_{\k+\q \sigma},~~ S^\alpha_{\q}=\frac{1}{2}\sum_{\k \sigma \sigma'} \sigma^\alpha_{\sigma \sigma'}c^\dagger_{\k \sigma} c_{\k+\q \sigma'},
\ee 
where $\sigma^\alpha$ is the $\alpha$ component of the Pauli matrix $\lp \alpha=x,y,z\rp$. As already mentioned, the superconducting gap for NCS superconductor is a mixture of singlet and triplet gap 
functions and is defined in the usual manner by the equation,
\be
\hat{\Delta}_\k=\lb\Psi(\k)\hat{\bf{\sigma}}_0+{\bf d(\k)}.\hat{\sigma}\rb i\hat{\sigma}_y.
\label{gap}
\ee
Here $\Psi(\k)$ is the singlet gap function and triplet $\bf d$-vector is ${\bf{d}}(\k)=(d_x(\k), d_y(\k), d_z(\k))$. In NCS superconductor triplet component satisfying 
$|{\bf d(\k)}.{\bf g}_{\k}|=|{\bf d(\k)}||{\bf g}_{\k}|$ only survives the pinning from the ASO coupling \cite{Sigrist}. So one can write $\bf d(\k)=\phi(k){\bf g}_{\k}$ where $\phi(\k)$ having the same 
symmetry of the momentum dependence as $\Psi(\k)$. For future reference we would like to mention that the static susceptibilities $\chi_{\alpha \beta}(\q)$ can be transformed to 
$\chi_{\sigma_1 \sigma_2 \sigma_3 \sigma_4}(\q)$ as,

\bea
\left(\begin{array}{rrrr}
\chi_{ss}(\q) \chi_{sz}(\q) \chi_{sx}(\q) \chi_{sy}(\q) \\
\chi_{zs}(\q) \chi_{zz}(\q) \chi_{zx}(\q) \chi_{zy}(\q) \\
\chi_{xs}(\q) \chi_{xz}(\q) \chi_{xx}(\q) \chi_{xy}(\q) \\
\chi_{ys}(\q) \chi_{yz}(\q) \chi_{yx}(\q) \chi_{yy}(\q) \end{array} \right)=\frac{1}{2} \hat W \left(\begin{array}{rrrr}
\chi_{\uparrow \uparrow \uparrow \uparrow}(\q) \chi_{\uparrow \uparrow \downarrow \downarrow}(\q) \chi_{\uparrow \uparrow \uparrow \downarrow}(\q) \chi_{\uparrow \uparrow \downarrow \uparrow }(\q)\\
\chi_{\downarrow \downarrow \uparrow \uparrow}(\q) \chi_{\downarrow \downarrow \downarrow \downarrow}(\q) \chi_{\downarrow \downarrow \uparrow \downarrow}(\q)\chi_{\downarrow \downarrow \downarrow \uparrow}(\q) \\
\chi_{\uparrow \downarrow \uparrow \uparrow}(\q) \chi_{\uparrow \downarrow \downarrow \downarrow}(\q) \chi_{\uparrow \downarrow \uparrow \downarrow}(\q)\chi_{\uparrow \downarrow \downarrow \uparrow}(\q)\\
\chi_{\downarrow \uparrow \uparrow \uparrow}(\q) \chi_{\downarrow \uparrow \downarrow \downarrow}(\q) \chi_{\downarrow \uparrow \uparrow \downarrow}(\q)\chi_{\downarrow \uparrow \downarrow \uparrow}(\q) 
\end{array} \right) \hat W^ \dagger
\label{chimat}
\eea
Here $\hat W $ is the orthogonal matrix of transformation given as,
\bea
\hat W = \left( \begin{array}{rrrr}
1 & 1 & 0 & 0 \\
1 & -1 & 0 & 0 \\
0 & 0 & 1 & i \\
0 & 0 & 1 & -i \end{array} \right),
\eea
and $\chi_{\sigma_1 \sigma_2 \sigma_3 \sigma_4}(\q)$ is defined as,
\be
\chi_{\sigma_1 \sigma_2 \sigma_3 \sigma_4}(\q)= -T \sum_{\k} \sum_{m} G_{\sigma_3 \sigma_1}(\k,i\omega_m) G_{\sigma_2 \sigma_4}(\k+\q,i\omega_m).
\ee
Here $G_{\sigma_i \sigma_j}$ are the components of the Fourier transform of matrix normal Green's function defined by,
\be
G_{\sigma \sigma'}(\k,\tau)=-\langle T_\tau\lb c_{\k\sigma}(\tau)c_{\k\sigma'}^\dagger(0)\rb\rangle.
\label{normalG}
\ee
Similarly we can define components of matrix anomalous Green's function as,
\be
\F_{\sigma \sigma'}(\k,\tau)=\langle T_\tau\lb c_{\k\sigma}(\tau)c_{-\k\sigma'}(0)\rb\rangle.
\label{F}
\ee
Now we expand normal Green's function according to second order perturbation theory treating the Coulomb interaction part of $H$ in Eq.(\ref{H}) as a perturbation. Thus we have to evaluate the imaginary 
time-ordered bracket below by applying Wick's theorem,
\be
G_{\sigma \sigma'}(\k,\tau)=-\langle T_\tau\lb c_{\k\sigma}(\tau)U(\tau_1)U(\tau_2)c_{\k\sigma'}^\dagger(0)\rb\rangle,
\ee 
here $U(\tau)$ describe the Coulomb interaction part of H and have the general form,
\be
U(\tau)=U\sum_{\k\k'\q}\sum_{\sigma}c_{\k+\q\sigma}^\dagger(\tau)  c_{\k\sigma}(\tau) c_{\k'-\q \bar{\sigma}}^\dagger(\tau) c_{\k' \bar{\sigma}}(\tau).
\ee
After following the above mentioned steps we can arrive at the following expression for the order parameter \cite{Soumya},
\be
\Delta_{\sigma_1 \sigma_2}=\chi_{\bar{\sigma_1}\bar{\sigma_1}\bar{\sigma_2}\bar{\sigma_2}}\F_{\sigma_1 \sigma_2}-\chi_{\bar{\sigma_1}\sigma_1\bar{\sigma_2}\bar{\sigma_2}}\F_{\bar{\sigma_1} \sigma_2}
-\chi_{\bar{\sigma_1}\bar{\sigma_1}\sigma_2\bar{\sigma_2}}\F_{\sigma_1 \bar{\sigma_2}}+\chi_{\bar{\sigma_1}\sigma_1\sigma_2\bar{\sigma_2}}\F_{\bar{\sigma_1} \bar{\sigma_2}}.
\label{gap_app}
\ee
The form of the Feynman diagrams that contribute to the calculation of the self energy part of the Green's function are also listed in ref.\cite{Soumya}. We can also write the matrix anomalous Green's 
function Eq.(\ref{F}) in spin space as the combination of terms from singlet ($\F_s(k)$) and triplet part ($\F_{\alpha}(k), \alpha=x,y,z$) as shown below,
\be
\hat{\F}(\k)=\lb\F_s(k)\hat{\sigma_0}+\F_{\alpha}(k).\hat{\sigma}\rb i \hat{\sigma}_y.
\label{FST}
\ee
Using Eqs.(\ref{gap}), (\ref{chimat}), (\ref{gap_app}) and (\ref{FST}) we can evaluate the different contributions (singlet and triplet) to the superconducting gap function. The singlet part can be written 
as,
\bea
&&\psi_s= \frac{1}{2}(\Delta_{\uparrow\downarrow}-\Delta_{\downarrow\uparrow})= \nonumber \\ &&U^2(\chi_{ss}-\chi_{xx}-\chi_{yy}-\chi_{zz}) \F_s + U^2\lp (\chi_{sx}-\chi_{xs})+ i (\chi_{yz}-\chi_{zy})\rp \F_x
+\nonumber \\ \nonumber &&U^2 \lp (\chi_{sy}-\chi_{ys})+ i (\chi_{zx}-\chi_{xz})\rp \F_y + U^2\lp (\chi_{sz}-\chi_{zs})+ i (\chi_{xy}-\chi_{yx})\rp \F_z .
\eea
Since we are working with static susceptibilities so terms like $\chi_{cx}$ do not contribute. So we can finally write, 
\be
\psi_s= v_{ss}\F_s + v_{sx}\F_x + v_{sy}\F_y + v_{sz}\F_z, 
\ee
where 
\bea
&&v_{ss}= U^2(\chi_{ss}-\chi_{xx}-\chi_{yy}-\chi_{zz}), v_{sx}= iU^2 (\chi_{yz}-\chi_{zy}), v_{sy}= iU^2 (\chi_{zx}-\chi_{xz}),\nonumber \\ &&v_{sz}= iU^2 (\chi_{xy}-\chi_{yx}).
\eea
Similarly, we can evaluate the different components of the triplet $\bf{d}$-vector. We start with $d_x$, which is found to be,
\bea
&&d_x= \frac{1}{2}(\Delta_{\downarrow\downarrow}-\Delta_{\uparrow\uparrow})= \nonumber \\
&&U^2\lp (\chi_{sx}-\chi_{xs}) -i (\chi_{yz}-\chi_{zy})\rp \F_s + U^2(\chi_{ss}-\chi_{xx}+\chi_{yy}+\chi_{zz}) \F_x +  \nonumber
\\ \nonumber &&U^2\lp i(\chi_{sz}+\chi_{zs})-(\chi_{xy}+\chi_{yx})\rp \F_y - U^2\lp i(\chi_{sy}+\chi_{ys})+ (\chi_{zx}+\chi_{xz})\rp \F_z.
\eea
Which gives us, 
\be
d_x= v_{xs}\F_s + v_{xx}\F_x + v_{xy}\F_y + v_{xz}\F_z, 
\ee
where, 
\bea
&&v_{xs}= -iU^2(\chi_{yz}-\chi_{zy}), v_{xx}= U^2(\chi_{ss}-\chi_{xx}+\chi_{yy}+\chi_{zz}), v_{xy}= -U^2(\chi_{xy}+\chi_{yx}),\nonumber \\ && v_{xz}= -U^2(\chi_{zx}+\chi_{xz}).
\eea
We can also find out the expressions for both $d_y$ and $d_z$.
\be
d_z= \frac{1}{2}(\Delta_{\uparrow\downarrow}+\Delta_{\downarrow\uparrow})= v_{zs}\F_s + v_{zx}\F_x + v_{zy}\F_y + v_{zz}\F_z,
\ee
where, 
\bea
&&v_{zs}= -iU^2(\chi_{xy}-\chi_{yx}), v_{zx}= -U^2(\chi_{xz}+\chi_{zx}), v_{zy}= -U^2(\chi_{yz}+\chi_{zy}),\nonumber \\ &&v_{zz}= U^2(\chi_{ss}+\chi_{xx}+\chi_{yy}-\chi_{zz}).
\eea
Finally we give relevant expressions for $d_y$.
\be
d_y= -\frac{i}{2}(\Delta_{\downarrow\downarrow}+\Delta_{\uparrow\uparrow})= v_{ys}\F_s + v_{yx}\F_x + v_{yy}\F_y + v_{yz}\F_z,
\ee
where, 
\bea
&&v_{ys}= iU^2(\chi_{xz}-\chi_{zx}), v_{yx}= -U^2(\chi_{xy}+\chi_{yx}), v_{yy}= U^2(\chi_{ss}+\chi_{xx}-\chi_{yy}+\chi_{zz}),\nonumber \\ &&v_{yz}= -U^2(\chi_{yz}+\chi_{zy}).
\eea
Using the above expressions for $\chi_s, d_x, d_y$ and $d_z$ we can arrive at the superconducting gap equation (Eq.[14] of Ref.\cite{Soumya}).

Thus in this article we have derived the superconducting gap equations relevant for the NCS superconductor $Li_2Pt_3B$ about which we mentioned in Ref.\cite{Soumya}. This gap equation is finally solved and 
this gives rise to a $s_\pm$ type of gap function with line nodes. The appearance of these nodal lines is purely accidental and can not be explained following its point group symmetry. It particularly 
depends on the structure of the underlying Fermi surface. This particular form of the order parameter explains the line node behavior of the gap function observed in experiments. Further experimental as well 
as theoretical investigations are required to reach a conclusion regarding the symmetry of the gap function of the noncentrosymmetric superconductor $Li_2Pt_3B$.

\section*{Acknowledgments}
We acknowledge the Max Planck Society (MPG), the Korea Ministry of Education, Science and Technology (MEST), Gyeongsangbuk-Do and Pohang City for the support of the Independent Junior Research Groups at the 
Asia Pacific Center for Theoretical Physics (APCTP). T.T acknowledges the support by the National Research Foundation of Korea(NRF) funded by the Ministry of Education, Science and Technology 
(Grant No. 2012R1A1A2008559).

\end{document}